\documentstyle[prl,aps,preprint,floats,epsfig]{revtex}

\begin{document}

\title{Electromagnetic energy-momentum and forces in matter}

\author{Yuri N.\ Obukhov\footnote{On leave from: Department 
of Theoretical Physics, Moscow State University, 117234 Moscow, Russia}}
\address
{Instituto de Fisica Teorica, Universidade Estadual Paulista,
Rua Pamplona 145\\ 01405-900 S\~ao Paulo, Brazil,\\
{\rm and} Institute for Theoretical Physics, University of Cologne,
50923 K\"oln, Germany}
\author{Friedrich W. Hehl}
\address{Institute for Theoretical Physics, University of Cologne,
50923 K\"oln, Germany\\
{\rm and} Department of Physics and Astronomy,
University of Missouri-Columbia\\ Columbia, MO 65211, USA}

\maketitle

\begin{abstract}
  We discuss the electromagnetic energy-momentum distribution and the
  mechanical forces of the electromagnetic field in material media.
  There is a long-standing controversy on these notions. The Minkowski
  and the Abraham energy-momentum tensors are the most well-known ones. 
  We propose a solution of this problem which appears to be natural and
  self-consistent from both a theoretical and an experimental point
  of view. 
\end{abstract}




\section{Introduction}

The discussion of the energy-momentum tensor in macroscopic electrodynamics 
is quite old. The beginning of this dispute goes back to Minkowski 
\cite{minkowski}, Abraham \cite{abraham}, and Einstein and Laub \cite{laub}. 
Good reviews of the problem can be found in \cite{robin,brevik,skob,ginz},
to mention but a few. Nevertheless, up to now the question was not
settled and there is an on-going exchange of conflicting opinions 
concerning the validity of the Minkowski versus the Abraham energy-momentum
tensor, see \cite{antoci} for a recent discussion. Even experiments were
not quite able to make a definite and decisive choice of electromagnetic 
energy and momentum in material media. 

Here we propose the solution of the problem.

Our basic notations and conventions are as follows. We are using 
international SI units throughout. Correspondingly, $\varepsilon_0, \mu_0$ 
are the electric and the magnetic constant (earlier called vacuum permittivity 
and vacuum permeability).  The Minkowski metric is $g_{ij}={\rm diag}(c^2,
-1, -1, -1)$. Latin indices from the middle of the alphabet label the 
spacetime components, $i,j,k,\dots=0,1,2,3$, whereas those from the beginning 
of the alphabet refer to 3-space: $a,b,c,\dots=1,2,3$.

\section{New energy-momentum tensor}

Our solution is as follows. 
The new electromagnetic energy-momentum in an {\em arbitrary medium} reads:
\begin{equation}\label{TEM}
\mu_0\,T_i{}^j =-\,F_{ik}\,F^{jk} + {\frac 1 4}\,\delta_i^j\,F_{kl}F^{kl}.
\end{equation}
The electromagnetic field strength $F_{ij} = ({\bf E}, {\bf B})$ is
composed of the electric and magnetic 3-vector fields. Componentwise, 
Eq.(\ref{TEM}) describes the {\it energy density} of the field
\begin{equation}
T_0{}^0 = u = {\frac 1 2}\left(\varepsilon_0\,E^2 
+ {\frac 1 {\mu_0}}\,B^2\right),\label{uTEM}
\end{equation}
its {\it energy flux density} (or Poynting vector)
\begin{equation}\label{sTEM}
T_0{}^a = s^a = {\frac 1 {\mu_0}}\,\left[{\bf E}\times{\bf B}\right]^a,
\end{equation}
its {\it field momentum density} 
\begin{equation}\label{pTEM}
T_a{}^0 = -\,p_a = \varepsilon_0\,\left[{\bf B}\times{\bf E}\right]_a,
\end{equation}
and the Maxwell {\it stress} tensor
\begin{eqnarray}
T_a{}^b = S_a{}^b &=& \varepsilon_0\left(E_aE^b - {\frac 1 2}
\,\delta_a^b\,E^2\right)\nonumber\\
&&\quad + {\frac 1 {\mu_0}}\left(B_aB^b - {\frac 1 2}
\,\delta_a^b\,B^2\right).\label{StTEM}
\end{eqnarray}

As one can immediately notice that (\ref{TEM}) has the same {\it form} as 
the vacuum energy-momentum tensor. However, its physical {\it content} is 
different, which follows from the fact that $F_{ij} = ({\bf E}, {\bf B})$ 
satisfies the {\it macroscopic} Maxwell equations in matter:
\begin{eqnarray}
\nabla\times{\bf H} - \dot{\bf D} &=& {\bf j}^{\rm free},\qquad
\nabla\cdot{\bf D} = \rho^{\rm free},\label{inMax}\\
\nabla\times{\bf E} + \dot{\bf B} &=& 0,\qquad
\nabla\cdot{\bf B} = 0.\label{homMax}
\end{eqnarray}
Here $\rho^{\rm free}$ and ${\bf j}^{\rm free}$ are the densities of
the free (or, in other terminology, the ``true'' or ``external'')
charge and the free current. The fields ${\bf D}$ and ${\bf H}$
represent the electric and magnetic excitations (other names are
``electric displacement'' and ``magnetic field intensity''). The field
strengths $({\bf E}, {\bf B})$ and the excitations $({\bf D}, {\bf
  H})$ are related by means of the equations
\begin{equation}
\varepsilon_0\,{\bf E} = {\bf D} - {\bf P},\quad {\rm and}\quad
{\frac 1 {\mu_0}}\,{\bf B} = {\bf H} + {\bf M}.\label{EDPBHM}
\end{equation}
The polarization ${\bf P}$ and the magnetization ${\bf M}$ represent
the {\it bound} (or ``polarizational'') current and charge 
densities inside the material medium:
\begin{equation}
{\bf j}^{\rm mat} = \dot{\bf P} + \nabla\times{\bf M},\qquad
\rho^{\rm mat} = -\,\nabla\cdot{\bf P}.\label{defPM}
\end{equation}

The density of the mechanical (ponderomotive) force acting on matter 
is determined as the  divergence of the energy-momentum tensor
\begin{equation}
f_i = \partial_j\,T_i{}^j.\label{divT}
\end{equation}
Differentiating (\ref{TEM}) and using the Maxwell equations (\ref{inMax}), 
(\ref{homMax}) and the equations (\ref{EDPBHM}) and (\ref{defPM}), yields:
\begin{eqnarray}\label{f0tot}
f_0 &=&\dot{u}+\nabla\cdot{\bf s} = -\,{\bf E}\cdot{\bf j}^{\rm tot},\\
f_a &=& -\,\dot{p}_a + \partial_b\,S_a{}^b = \rho^{\rm tot}\,E_a 
+ \left[{\bf j}^{\rm tot}\times{\bf B}\right]_a.\label{ftot}
\end{eqnarray}
Here the total charge and current density are $\rho^{\rm tot} = 
\rho^{\rm free} + \rho^{\rm mat}$ and ${\bf j}^{\rm tot}=
{\bf j}^{\rm free} + {\bf j}^{\rm mat}$.

This result is quite natural and physically clear. The electromagnetic
field affects matter by means of the two Lorentz forces (\ref{ftot}):
One acts on the free charge and current $(\rho^{\rm free}, {\bf
  j}^{\rm free})$ (on the conductive current, for example), and
another force acts on the bound charge and current (\ref{defPM}). The
latter have also a direct physical meaning in the macroscopic (Lorentz
type averaging) framework and in the microscopic approaches, see Hirst
\cite{hirst}, e.g. The temporally and spatially varying polarization
and magnetization give rise to the electric and magnetic fields, like
the free charges and currents do. Conversely, the bound charges and
currents should also feel the electromagnetic field in the same way as
the free charges and currents do.

The representation of the total electromagnetic force as the sum of the two 
terms with the clear-cut physical content (\ref{ftot}) suggests a natural 
step to split the original energy-momentum (\ref{TEM}) into the corresponding
sum of the two energy-momentum tensors which are associated with the
free and bound charge/current, respectively. Using (\ref{EDPBHM}) in
(\ref{TEM}), we find
\begin{equation}
T_i{}^j = {}^{\rm f}T_i{}^j + {}^{\rm b}T_i{}^j,\label{sum}
\end{equation}
where we introduce the {\it free-charge} energy-momentum and the 
{\it bound-charge} energy-momentum tensors as 
\begin{equation}
{}^{\rm f}T_i{}^j = \left(\begin{array}{c|c} {}^{\rm f}u& -\,{}^{\rm f}p_a
\\ \hline {}^{\rm f}s^b & {}^{\rm f}S_a{}^b\end{array}\right),\qquad
{}^{\rm b}T_i{}^j = \left(\begin{array}{c|c} {}^{\rm b}u& -\,{}^{\rm b}p_a
\\ \hline {}^{\rm b}s^b & {}^{\rm b}S_a{}^b\end{array}\right).
\end{equation}
The components of the free-charge energy-momentum read explicitly
\begin{eqnarray}
{}^{\rm f}u &=& {\frac 1 2}\left({\bf E}\cdot{\bf D} 
+{\bf B}\cdot{\bf H}\right),\label{uMink}\\
{}^{\rm f}{\bf s} &=& {\bf E}\times{\bf H},\qquad 
{}^{\rm f}{\bf p} = {\bf D}\times{\bf B},\label{pMink}\\
{}^{\rm f}S_a{}^b &=& E_aD^b + B_aH^b - {\frac 1 2}\,\delta_a^b
\left({\bf E}\cdot{\bf D} +{\bf B}\cdot{\bf H}\right),\label{SMink}
\end{eqnarray}
whereas the components of the bound-charge energy-momentum are
\begin{eqnarray}
{}^{\rm b}u &=& {\frac 1 2}\left(-\,{\bf E}\cdot{\bf P} 
+{\bf B}\cdot{\bf M}\right),\label{ubound}\\
{}^{\rm b}{\bf s} &=& {\bf E}\times{\bf M},\qquad 
{}^{\rm b}{\bf p} = -\,{\bf P}\times{\bf B},\label{pbound}\\
{}^{\rm b}S_a{}^b &=& -\,E_aP^b + B_aM^b - {\frac 1 2}\,\delta_a^b
\left(-\,{\bf E}\cdot{\bf P} +{\bf B}\cdot{\bf M}\right).\label{Sbound}
\end{eqnarray}

One straightforwardly recognizes the tensor ${}^{\rm f}T_i{}^j$ with 
the components (\ref{uMink})-(\ref{SMink}) as the well-known Minkowski 
energy-momentum tensor. 

Similarly to (\ref{divT}), the  divergences ${}^{\rm f}f_i 
= \partial_j{}^{\rm f}T_i{}^j$ and ${}^{\rm b}f_i = \partial_j
{}^{\rm b}T_i{}^j$ determine the force densities. By construction, the 
total 4-force density is the sum $f_i = {}^{\rm f}f_i + {}^{\rm b}f_i$. 
Explicitly, we have for the 3-force densities
\begin{eqnarray}
{}^{\rm f}f_a &=& \rho^{\rm free}\,E_a 
+ \left[{\bf j}^{\rm free}\times{\bf B}\right]_a - X_a,\label{3fM}\\
{}^{\rm b}f_a &=& \rho^{\rm mat}\,E_a 
+ \left[{\bf j}^{\rm mat}\times{\bf B}\right]_a + X_a.\label{3fB}
\end{eqnarray}
Here, in the Cartesian coordinates,
\begin{equation}
X_a = {\frac 12}\left(E^b\partial_a D_b - D^b\partial_a E_b + 
H^b\partial_a B_b - B^b\partial_a H_b\right). 
\end{equation}
In particular, $X_a =\!{\frac 12}\!\left[\varepsilon_0\,E_bE_c\partial_a
\varepsilon^{bc} + \mu_0\,H_bH_c\partial_a\mu^{bc}\right]$ for the 
{\it linear} material laws $D^a = \varepsilon_0\,\varepsilon^{ab}E_b$ and 
$B^a =\mu_0\,\mu^{ab}H_b$. This extra term vanishes for homogeneous media.

Thus indeed, tensors in the sum (\ref{sum}) are associated with the two 
different types of charges and currents in the material medium. The 
 divergences produce, essentially, the two independent Lorentz 
forces acting separately on the free and on the bound charge and current.

\section{Some properties and applications}

After all these preliminaries and formal derivations, we are in a position
to discuss the physical properties of the energy-momentum (\ref{TEM}). At
first, some remarks about (\ref{TEM}) in comparison to the Minkowski and 
Abraham tensors. Many authors (see the discussions in \cite{brevik,skob,ginz}) 
pointed to a clearly unphysical result produced by the Minkowski 
energy-momentum: in the absence of free charges and currents, 
a homogeneous medium appears to be always subject to the {\it zero} 
electromanetic force. This fact was usually taken in favor of the Abraham
tensor which predicts an extra, so called Abraham force. However, the
energy-momentum (\ref{TEM}) does not suffer from such a deficiency. Even
when the free charge and current densities are vanishing, the total force is, 
in general, non-trivial in view of the presence of polarization charge and 
current. Moreover, as compared to the rather ad hoc choice of the Abraham 
force, the mechanical action on the bound charge and current is in all 
cases described, using (\ref{TEM}), by the well-known Lorentz force 
(\ref{ftot}). 

Furthermore, the Minkowski tensor is asymmetric which is obvious from
the comparison of the energy flux ${}^{\rm f}{\bf s}$ and the field
momentum ${}^{\rm f}{\bf p}$ in (\ref{pMink}). Usually, this fact was
also taken in favor of the Abraham tensor, which is symmetric. At the
same time, despite its symmetry, the structure of the Abraham tensor
is defined in a rather ad hoc manner with opaque physical motivations.
In contrast to this, the energy-momentum (\ref{TEM}) is naturally
symmetric and the electromagnetic field momentum (\ref{pTEM}) is
related to the energy flux (\ref{sTEM}) as ${\bf p} = {\bf s}/c^2$.
This is the famous Planck relation which generalizes the Einsteinian
$\Delta m = \Delta E/c^2$ relation to field theory. The interesting 
discussion of von Laue \cite{laue} was concentrated mainly around 
this point.

As a simple application, let us consider the propagation of an
electromagnetic plane wave from the vacuum into a dielectric medium
with $\mu=1$ and refractive index $n=\sqrt{\varepsilon}$. More exactly,
like in the previous discussions \cite{gordon,skob,ginz}, we will confine
ourselves to the case of a gaseous medium consisting of heavy atoms. We 
assume normal incidence on a plane boundary and we recall the reflection and
transmission coefficients $R=(n-1)/(n+1)$ and $T=2/(n+1)$, respectively. 
Then, for incident and reflected waves in vacuum, we find for the mean 
field energy (\ref{uMink}) and momentum (\ref{pMink}), if averaged over 
one period,
\begin{eqnarray}
\overline{u} &=& \varepsilon_0\,{\frac {|E_0|^2} 2}\,(1 + R^2)
= \varepsilon_0\,|E_0|^2\,{\frac {1+n^2}{(1+n)^2}},\\
\overline{\bf p} &=& {\frac {\varepsilon_0} c}\,{\frac {|E_0|^2} 2}
\,(1 - R^2)\,{\bf k} = {\frac {\varepsilon_0} c}\,|E_0|^2
\,{\frac {2n}{(1+n)^2}}\,{\bf k}.
\end{eqnarray}
On the other hand, within the dielectric, for the transmitted wave, Eqs.
(\ref{uMink}), (\ref{ubound}) and (\ref{pMink}), (\ref{pbound}) yield
\begin{eqnarray}
\overline{u} &=& \varepsilon_0\,{\frac {1 + n^2} 2}\,{\frac {|E_0|^2} 2}
\,T^2 = \varepsilon_0\,|E_0|^2\,{\frac {1+n^2}{(1+n)^2}},\\
\overline{\bf p} &=& {\frac {\varepsilon_0n} c}\,{\frac {|E_0|^2} 2}
\,T^2\,{\bf k} = {\frac {\varepsilon_0} c}\,|E_0|^2
\,{\frac {2n}{(1+n)^2}}\,{\bf k}.
\end{eqnarray}
Here $E_0$ is the amplitude of the electric field and ${\bf k}$ the unit 
wave vector which specifies the direction of propagation. Comparing the 
above formulas, we see that both, the total energy and the total momentum 
calculated on the basis of (\ref{TEM}), are conserved on the passage of 
the wave into the medium. This conclusion is in a complete agreement 
with the previous studies \cite{gordon,skob,ginz}. Moreover, it can be 
supplemented by a far more detailed analysis of wave propagation in 
a gaseous media that has demonstrated \cite{gordon,skob,ginz} the 
plausibility of the (``Abraham'') field momentum ${\bf E}\times{\bf H}/c$. 
However, since all these studies were confined to dielectrics with 
$\mu=1$, the arguments presented in \cite{gordon,skob,ginz} in actual fact 
give support to the field momentum (\ref{pTEM}) likewise.

\section{Walker and Walker experiment}

Finally, let us discuss the direct {\it experimental confirmations} of the
energy-momentum (\ref{TEM}). For this purpose, as a first example, we
recall the measurements of Walker \& Walker \cite{walker} of the force
acting on a dielectric disk placed in crossed oscillating electric and
magnetic fields, see Fig.1(a). The scheme of the experiment is as
follows: The symmetry of the problem suggests to use cylindrical
coordinates $(r, \varphi, z)$.  A small cylinder is made of barium
titanate with $\varepsilon = 3340$ and $\mu=1$. Its height in
$z$-direction is $l\approx 2$ cm, internal radius $r_1\approx 0.4$ cm,
and external radius $r_2\approx 2.6$ cm. This disk is suspended
between the poles of an electromagnet which creates the harmonically
oscillating {\it axial} magnetic field. Besides, the oscillating
field, with the phase difference of $\pi/2$, a {\it radial} electric
field is created by means of an alternating voltage applied between
the inner and the outer cylindrical surfaces of the disk. The oscillation
frequency of the fields is rather low, namely, $\omega = 60$~Hz.  As a
consequence, everywhere in the disk $r\omega n/c\sim z\omega n/c\sim
10^{-7}\ll 1$.  Walker \& Walker \cite{walker} measured the torque around 
the $z$-axis produced by the electromagnetic force.

Let us derive the theoretical value of the torque which is given by
the volume integral 
\begin{equation}\label{torque1}
N_z = \int_{\rm disk}dv\,[{\bf r}\times{\bf f}]_z = 
\int_{\rm disk}dv\,rf_\varphi.
\end{equation}
Since there are no free charges and currents in the dielectric, the
Lorentz force (\ref{ftot}) reduces only to the second term determined
by the bound charges and currents. One can check that the Maxwell equations
(\ref{inMax}), (\ref{homMax}), together with the constitutive relations
${\bf D}=\varepsilon\varepsilon_0{\bf E}$ and ${\bf B}=\mu_0{\bf H}$, are 
solved by the electric and magnetic field configuration 
\begin{eqnarray}
{\bf E} &=& \sin(\omega t)\left({\frac {U_0}{r\log(r_2/r_1)}}
\,{\bf e}_r + {\frac {B_0} 2}\,\omega r\,{\bf e}_\varphi\right),\label{E}\\
{\bf B} &=& \cos(\omega t)\left(B_0\,{\bf e}_z - 
{\frac {\varepsilon U_0}{r\log(r_2/r_1)\,c^2}}
\,\omega z\,{\bf e}_\varphi\right).\label{B}
\end{eqnarray}
These approximate formulas are valid with very high accuracy due to 
the fact that $r\omega n/c\sim z\omega n/c\sim 10^{-7}\ll 1$ everywhere 
in the disk. As usual, $({\bf e}_r, {\bf e}_\varphi, {\bf e}_z)$ denote 
the local orthonormal frame vectors of the cylindrical coordinate system.
Here $B_0$ is the magnitude of the oscillating axial magnetic field and
$U_0$ is the amplitude of the voltage applied between the inner ($r=r_1$) 
and the outer ($r=r_2$) cylindrical surfaces of the disk, $\Delta U = 
U_0\,\sin(\omega t)$.

The bound charge and current densities (\ref{defPM}) are straightforwardly 
found to be $\rho^{\rm mat} =0$ and ${\bf j}^{\rm mat} = \dot{\bf P} =
\varepsilon_0(\varepsilon -1)\dot{\bf E}$. Correspondingly, the Lorentz
force turns out to be ${\bf f} = {\bf j}^{\rm mat}\times{\bf B}=
\varepsilon_0(\varepsilon -1)\dot{\bf E}\times{\bf B}$. Substituting 
(\ref{E}) and (\ref{B}) in (\ref{torque1}), we obtain the torque
\begin{equation}\label{torque2}
N_z = -\,\varepsilon_0(\varepsilon - 1)\,\pi\,l\,(r_2^2 - r_1^2)
\,{\frac {U_0B_0\omega}{\log(r_2/r_1)}}\,\cos^2(\omega t).
\end{equation}
This result was experimentally confirmed by Walker \& Walker \cite{walker}.
Although the authors of \cite{walker} apparently noticed that the torque 
measured fits the understanding of the electromagnetic force as the 
Lorentz force for the polarization current (in agreement with our approach), 
they ultimately claimed that their experiment confirms the Abraham force. The 
theoretical explanation presented in \cite{walker} was based on the idea 
that the Maxwell stress $S_a{}^b$ caused the unusual surface drag of the 
disk. However, when computing the force, see (\ref{ftot}), the contribution 
of the stress $\partial_b\,S_a{}^b$ should be complemented by the momentum 
term $-\,\dot{p}_a$; the latter is missing in \cite{walker}. The alternative
explanation \cite{brevik} is based on the computation of forces on the
metal coatings of the cylinder. This yields the result where the factor
$(\varepsilon -1)$ above is replaced by $\varepsilon$. Since the dielectric
matter in question has $\varepsilon = 3340$, this experiment thus cannot
be treated as the critical test for the different energy-momenta.

\section{Experiment of James}

As another example we consider the experiment of James \cite{james}
which is in many respects very similar to the one of Walker
\& Walker. James, see Fig.1(b), had also placed a disk into
crossed electric and magnetic fields. The small cylinders were made
of a composition of nickel-zinc ferrite with $\mu = 16$ or $43$ and
$\varepsilon\approx 7$.  Like in \cite{walker}, the {\it
  radial} electric field was created by means of an oscillating voltage
applied between the inner and the outer cylindrical surface of the
disk.  However, instead of an axial magnetic field, an {\it azimuthal}
magnetic field was produced inside matter by an alternating electric
current in a conducting wire placed along the axis of the disk. The
resulting field configuration reads: 
\begin{eqnarray}
{\bf E} &=& \left({\frac {U_0\,\sin(\omega_u t)}{r\log(r_2/r_1)}} 
- {\frac {\mu\mu_0I_0}{2\pi r}}\,\cos(\omega_i t)\,\omega_i z
\right){\bf e}_r,\label{EJ}\\
{\bf B} &=& \left({\frac {\mu\mu_0I_0}{2\pi r}}\,\sin(\omega_i t)
- {\frac {\mu\varepsilon\,U_0\,\cos(\omega_u t)}{r\log(r_2/r_1)\,c^2}}
\,\omega_u z\right){\bf e}_\varphi.\label{BJ}
\end{eqnarray}
These formulas hold true in the approximation $z\omega_u n/c\sim 
z\omega_i n/c \sim 10^{-5} \ll 1$ which is fulfilled in James' experiment 
everywhere in the cylinders with $r_1, r_2$ and the length $l$ of order
1-3 centimeter. Here $I_0$ is the amplitude of the alternating current $I=I_0
\,\sin(\omega_i t)$ along the $z$-axis, whereas $U_0$ gives, as before, the 
amplitude of the oscillating voltage, $\Delta U = U_0\,\sin(\omega_u t)$.
The frequencies $\omega_i$ and $\omega_u$ are different and are varied
in the course of the experiment between 10 and 30 kHz. Since this 
experiment, unlike \cite{walker}, covers also magnetic media, we
display the nontrivial permeability $\mu$. One can check that
(\ref{EJ}), (\ref{BJ}) satisfy the Maxwell equations (\ref{inMax}), 
(\ref{homMax}).

\subsection{Electromagnetic force in James' experiment}

James \cite{james} measured the force ${\cal F}_z$ acting along the axis 
of the disk in the crossed fields (\ref{EJ}), (\ref{BJ}). Let us derive 
the theoretical value of this force by using the general expression for 
the force density (\ref{ftot}). There are no free charges and currents 
inside matter, $\rho^{\rm free}=0$ and ${\bf j}^{\rm free}=0$. By substituting 
(\ref{EJ}), (\ref{BJ}) into (\ref{defPM}), we find $\rho^{\rm mat}=0$ and 
\begin{equation}
{\bf j}^{\rm mat} = \varepsilon_0\,(\mu\varepsilon - 1)\Bigg(
{\frac {\omega_u\,U_0\,\cos(\omega_u t)}{r\log(r_2/r_1)}}
+ {\frac {\mu\mu_0I_0}{2\pi r}}\,\sin(\omega_i t)
\,\omega_i^2 z \Bigg){\bf e}_r.\label{currJ}
\end{equation}
The total force is obtained as the integral of the force density 
\begin{eqnarray}
{\bf j}^{\rm mat}\times{\bf B} &=& {\frac {(\varepsilon\mu -1)\varepsilon_0} 
{r^2}}\Big[ {\frac {\mu\mu_0\,U_0I_0} {2\pi\log(r_2/r_1)}}
\,\omega_u\,\sin(\omega_i t)\cos(\omega_u t)\nonumber\\ && 
-\,\left(nU_0/c\log(r_2/r_1)\right)^2\,\omega_u^2\,\cos^2(
\omega_u t)\,z + (\mu\mu_0I_0/2\pi)^2\,\omega_i^2\,\sin^2(\omega_i t)
\,z\Big]\,{\bf e}_z
\end{eqnarray}
over the volume of the disk:
\begin{equation}
{\cal F}_z = \int_{\rm disk}dv\,[{\bf j}^{\rm mat}\times{\bf B}]_z 
= {\frac {(\mu\varepsilon - 1)\mu l U_0I_0} {c^2}}\,\omega_u\,\sin(\omega_i t)
\cos(\omega_u t).\label{forceJ0}
\end{equation}
According to James \cite{james}, we choose $\omega_i = \omega_u \pm \omega_0$,
with $\omega_0$ the mechanical resonance frequency of the disk, and find the 
final expression for the force
\begin{equation}
{\cal F}_z = \pm\,{\frac {(\mu\varepsilon - 1)
\mu l U_0I_0} {2c^2}}\,\omega_u\,\sin(\omega_0 t).\label{forceJ}
\end{equation}
This theoretical prediction was actually verified in the experiment of James 
\cite{james}.

\subsection{Minkowski and Abraham forces in James' experiment}

In the isotropic case under consideration, $\varepsilon^{ab} =
\delta^{ab} \,\varepsilon$ and $\mu^{ab} = \delta^{ab}\,\mu$. With the
free charges and currents absent, the Minkowski 3-force density
(\ref{3fM}) reduces to the last term which contributes only at the
ends of the cylinder. Since the permittivity has the
constant value of $\varepsilon\neq 1$ inside the body, i.e. for
$-l/2<z<l/2$, and drops to $\varepsilon = 1$ outside of that interval, the
derivative of such a stepwise function reads $\partial_z\varepsilon(z)
= (\varepsilon - 1)\left[\delta(z + l/2) - \delta(z - l/2)\right]$. 
Similar relation holds for the derivative of the permeability function
$\mu(z)$.  Correspondingly, we find for the Minkowski force
\begin{eqnarray}
{\cal F}_z^{\rm M} &=& -\,\int_{\rm disk} dv X_z = -\,{\frac 12}
\int_{\rm disk} dv\left[\varepsilon_0\,E^2\,\partial_z \varepsilon 
+ \mu_0\,H^2\,\partial_z\mu\right]\nonumber\\
&=& \pi \int_{r_1}^{r_2} dr\,r\left\{(\varepsilon - 1)\varepsilon_0
\left[ E^2(l/2) - E^2(-l/2)\right] + (\mu - 1)\mu_0\left[H^2(l/2) 
- H^2(-l/2)\right]\right\}.
\end{eqnarray}
Substituting the squares of the electric and magnetic fields (\ref{EJ}) 
and (\ref{BJ}) [note that ${\bf H} = {\bf B}/\mu\mu_0$], we obtain:
\begin{equation}
{\cal F}_z^{\rm M} = -\,{\frac {l\,U_0I_0}{c^2}}\left[\mu(\varepsilon -1)
\,\omega_i\,\sin(\omega_ut)\cos(\omega_it) + \varepsilon(\mu -1)\,\omega_u
\,\sin(\omega_it)\cos(\omega_ut)\right].\label{FM0}
\end{equation}

The Abraham 3-force density differs from the Minkowski expression by
the so-called Abraham term [see \cite{brevik}, eq. (1.6) on page 140,
for example]:
\begin{equation}
{\bf f}^{\rm A} = {\bf f}^{\rm M} + {\frac {\mu\varepsilon -1}{c^2}}
\,{\frac \partial {\partial t}}\,({\bf E}\times{\bf H}).\label{3fA}
\end{equation}
It is straightforward to evaluate the last term. Using (\ref{EJ}) and
(\ref{BJ}), we get
\begin{eqnarray}
{\frac 1 {c^2}}\,{\bf E}\times{\bf H} &=& {\frac {\varepsilon_0} {\mu r^2}}
\Big[{\frac {\mu\mu_0\,U_0I_0} {2\pi\log(r_2/r_1)}}\,\sin(\omega_ut)
\sin(\omega_it) - (\mu\mu_0I_0/2\pi)^2\,\omega_i\,\sin(\omega_it)
\cos(\omega_it)\,z \nonumber\\ 
&& -\,\left(nU_0/c\log(r_2/r_1)\right)^2\,\omega_u\,\sin(\omega_ut)
\cos(\omega_ut)\,z\Big]\,{\bf e}_z.
\end{eqnarray}
Taking the time derivative and integrating over the body, we find
an additional contribution to the total force:
\begin{eqnarray}
\Delta{\cal F}_z &=& \int_{\rm disk} dv\,{\frac {\mu\varepsilon -1}{c^2}}
\,{\frac \partial {\partial t}}\,({\bf E}\times{\bf H})_z\nonumber\\ 
&=& {\frac {(\mu\varepsilon -1)\,lU_0I_0}{c^2}}\left[\omega_i\,\sin(\omega_ut)
\cos(\omega_it) + \omega_u\,\sin(\omega_it)\cos(\omega_ut)\right].\label{FA0}
\end{eqnarray}

In James' experiment, we put $\omega_i = \omega_u \pm \omega_0$ and 
select only the component varying with the mechanical resonance frequency 
of the body, $\omega_0$. Then (\ref{FM0}) and (\ref{FA0}) yield the 
Minkowski and the Abraham forces:
\begin{eqnarray}
{\cal F}_z^{\rm M} &=& {\frac {l\,U_0I_0}{2c^2}}\left[\mu\,(\varepsilon -1)\,
\omega_0 \mp (\mu - \varepsilon)\,\omega_u\right]\sin(\omega_0t),\label{FM1}\\
{\cal F}_z^{\rm A} &=& {\cal F}_z^{\rm M} + \Delta{\cal F}_z = {\frac 
{l\,U_0I_0}{2c^2}}\left[(1 - \mu)\,\omega_0 \mp (\mu - \varepsilon)
\,\omega_u\right]\sin(\omega_0t).\label{FA1} 
\end{eqnarray}

\subsection{Theories versus experiment}

All the theoretical expressions for the electromagnetic force look similar:
compare (\ref{forceJ}) with (\ref{FM1}) and (\ref{FA1}). However, the 
crucial difference is revealed when we recall that James measured not 
the force itself but a ``reduced force" defined as the mean value 
${\frac 12}\left[ {\cal F}_z(\omega_u,\omega_i = \omega_u + \omega_0) 
+ {\cal F}_z(\omega_u, \omega_i = \omega_u - \omega_0)\right]$, see
eq. (9) on page 60 of James' thesis \cite{james} and the footnote on 
page 158 of \cite{brevik}. With high accuracy, James observed the 
{\it vanishing} of the reduced force in his experiment.
This observation is in complete agreement with the theoretical
derivation (\ref{forceJ}) based on our new energy-momentum tensor,
whereas both, the expressions of Minkowski (\ref{FM1}) and of Abraham
(\ref{FA1}), clearly contradict this experiment.

The explanation proposed in \cite{brevik} in support of the Abraham force 
appears to be inconsistent mathematically and misleading physically. 
Namely, the computation of the force is reduced in \cite{brevik} to the
evaluation of the surface integral of the Maxwell stress in the vacuum 
``just outside the disk''. However, instead of the usual continuity of the 
tangential electric field, an unsubstantiated matching condition 
was introduced for $\bf E$ on the boundary ($z=\pm l/2$) in order to find 
the fields outside the disk. 
Such a derivation [which yields a result different from (\ref{FA1}) above] 
cannot be considered to be a satisfactory theoretical explanation.

To begin with, there is not any good reason why one should replace a 
well-defined volume integral for the total force by a surface integral.  
Formally, this is allowed, of course, but as soon as we know the fields 
{\it inside\/} the body everywhere, see (\ref{EJ}) and (\ref{BJ}), we can 
proceed directly by constructing the explicit expressions of the force 
densities (\ref{3fM}), (\ref{3fB}), and (\ref{3fA}) and then 
straightforwardly find the corresponding volume integrals. 
There is no logical need to perform an auxiliary computation in order to find 
the vacuum fields ``just outside" the body, which appears to be a separate
nontrivial problem. Provided the latter problem is solved correctly, we 
anticipate that the final result would agree with our (\ref{FA1}). 
And certainly, one should use the standard matching conditions since 
this amounts to nothing else than to apply Maxwell's equations
in a thin neighborhood near the surface. Accordingly, as to the
matching of the electric field, we can only impose (as usual) the
{\it continuity\/} of the tangential components of electric field.
Imposing a different {\it discontinuity\/} condition [as was done in
eq. (3.17) of ref.  \cite{brevik}, e.g.] is tantamount to assuming
that Maxwell's equations are violated near and across the boundary.

In our theoretical analysis, we used the field configuration 
(\ref{EJ}), (\ref{BJ}) which is valid inside of the cylinder. Near the
ends, strictly speaking, one should take into account the deformation 
of the fields. However, it is well known that the corresponding corrections
are confined to the regions very close to the ends. More exactly, the most
important point is that the resulting end corrections for the total force 
are {\it not} proportional to the length of the cylinder. In other words,
such end corrections (provided one computes them carefully) obviously 
would not compensate the reduced force of Minkowski (\ref{FM1}) and 
of Abraham (\ref{FA1}), which are both proportional to the length $l$. 
It is worthwhile to note that the end corrections were never taken into 
account in the previous analyses \cite{james,brevik}, and we use here 
precisely  the same field configuration (\ref{EJ}), (\ref{BJ}) as in 
\cite{james,brevik}.

\section{Discussion and conclusion}

Let us summarize our results. In the present paper we gave evidence 
that the correct energy-momentum tensor of the electromagnetic field in
material media is described by (\ref{TEM}). This tensor is symmetric
and satisfies Planck's field-theoretical generalization ${\bf p} = 
{\bf s}/c^2$ of Einstein's formula $\Delta m = \Delta E/c^2$. The 
corresponding electromagnetic force turns out to be the Lorentz force 
acting on the free and bound charge and current densities. The 
energy-momentum (\ref{TEM}) can be naturally represented as a sum 
(\ref{sum}) of the Minkowski energy-momentum and the bound-charge 
energy-momentum tensor. 

Our derivations here are in fact motivated by our axiomatic approach to
classical electrodynamics \cite{book} in which the Lorentz force
represents one of the fundamental postulates of the scheme. In particular,
if one starts from the Lorentz force equations (\ref{f0tot}) and (\ref{ftot})
and reverses the order of the equations, one finally derives the 
energy-momentum tensor (\ref{TEM}) that we first introduced without 
preliminary explanations. Besides the evidence of the general validity of 
the Lorentz force axiom for point particles, a careful analysis of the wave 
propagation in material media as well as a proper interpretation of the 
experiments by Walker \& Walker and by James, give further support to this
basic cornerstone of classical electrodynamics. In our discussion we
did not touch the electro- and magnetostriction effects because their
consideration requires a more detailed specification of the internal
mechanical properties of the medium. Moreover, in most cases the overall
electro- and magnetostriction effects are balanced and are not
directly observable.

At the present level of understanding, we can thus conclude that the
tensor (\ref{TEM}) passes the theoretical and experimental tests and 
qualifies for a correct description of the energy-momentum properties
of the electromagnetic field in macroscopic electrodynamics. 

As we have learned recently, the same energy-momentum tensor was 
introduced by P.~Poincelot \cite{poin} who  insisted on the equal physical 
treatment of the free and the polarizational charges and currents. Such an
equality is natural in our axiomatic approach to electrodynamics \cite{book}.

\bigskip
{\bf Acknowledgments}. YNO's work was partially supported by FAPESP, 
and by the Deutsche Forschungsgemeinschaft (Bonn) with the grants 
436 RUS 17/70/01 and HE~528/20-1. 


\begin{figure}
\epsfxsize=\hsize \epsfbox{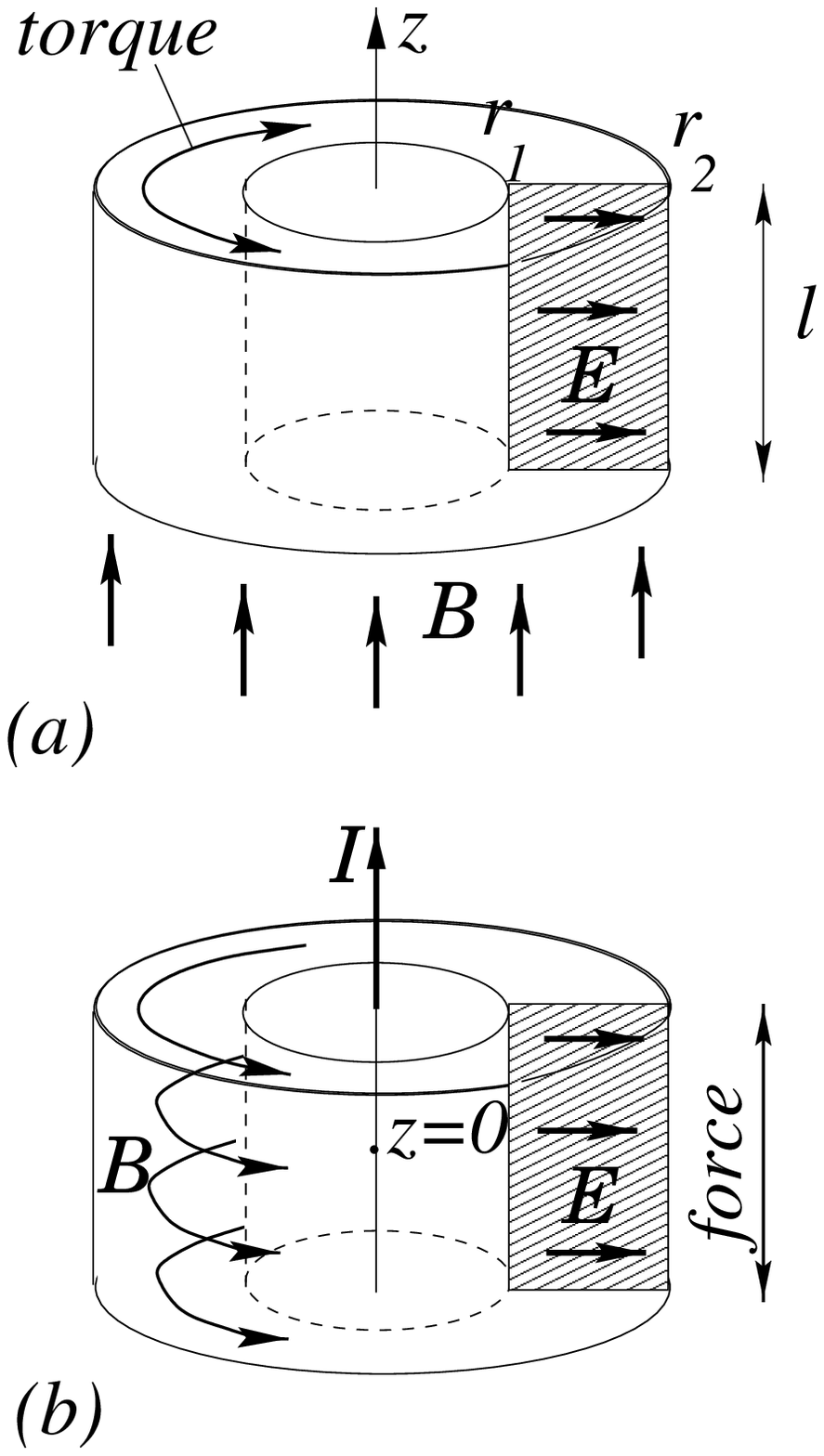}
\caption[(a) Walker \& Walker experiment; (b) James's experiment.]
        {(a) Walker \& Walker experiment; (b) James's experiment.}
\end{figure}

\end{document}